\documentclass[prl,aps,twocolumn, floatfix, superscriptaddress,showpacs]{revtex4}
\usepackage{amsmath,bm,graphicx}
\usepackage{amssymb}
\usepackage{amsfonts}
\begin{document}
\title{Local power fluctuations in two-dimensional turbulence.}
\author{Mahesh Bandi}
\email[Corresponding Author: ]{mbandi@lanl.gov}
\affiliation{CNLS \& MPA-10, Los Alamos National Laboratory, Los Alamos, NM 87545, USA}
\author{Colm Connaughton}
\email{connaughtonc@gmail.com}
\affiliation{CNLS \& T-13, Los Alamos National Laboratory, Los Alamos, NM 87545, USA}
\affiliation{Centre for Complexity Science \& Mathematics Institute, University of Warwick, Coventry CV4 7AL, UK}

\date{\today}
\begin{abstract}
The statistics of power fluctuations are studied in simulations of 
two-dimensional turbulence in both inverse (energy) and direct (enstrophy) cascade regimes from both Lagrangian and Eulerian perspectives. The probability density function (PDF) of the appropriately defined dimensionless power is strongly non-gaussian with asymmetric exponential tails. This distribution can be modeled by the distribution of the product of correlated normal variables allowing a derivation of the asymptotics of the tails. The PDF of the dimensionless power is shown to exhibit an empirical Fluctuation Relation. An expression for the entropy production rate is deduced from the asymptotic form of the power PDF and is found to agree very well with the measured entropy rate.
\end{abstract}
\maketitle

\section{Introduction}
Thermodynamics provides a theoretical framework for describing the macroscopic
properties of equilibrium systems which are largely independent of microscopic
details. Much research in recent years has searched for universal features in 
non-equilibrium systems to help develop macroscopic theories of 
non-equilibrium statistical physics that are similarly insensitive to
microscopic details. One important branch of this research has focused on
the Gallavotti--Cohen Fluctuation Theorem ~\cite{GC1995} and related results 
~\cite{JAR1997a,KUR1998,CRO1998} which are
collectively referred to as Fluctuation Relations. A Fluctuation Relation (FR)
is formulated for the PDF of some quantity, $X_\tau$,
derived from the entropy production or energy dissipation in a system  driven
far from equilibrium. $X_\tau$ is obtained by averaging a physical
quantity, $x(t)$, typically the entropy
produced or energy dissipated along a phase space trajectory of the system, over
a time interval $[t,t+\tau]$: $X_\tau = \tau^{-1}\ \int_t^{t+\tau} x(t^\prime)\ dt^\prime$.  
$X_\tau$ is positive on average but, may fluctuate 
sufficiently about its mean that negative fluctuations
are observable. An FR quantifies the relative
probability of observing a negative fluctuation over a given time interval
compared to the probability of observing a positive fluctuation of the same
magnitude. The ratio of probabilities takes the form:
\begin{equation}
\label{eq-generalFR}
\frac{\Pi(X_\tau)}{\Pi(-X_\tau)} = e^{\Sigma\,\tau\,X_\tau},
\end{equation}
where $\Sigma$ is a constant, independent of the averaging interval, $\tau$. 
In the original 
formulation of ~\cite{GC1995}, $x(t)$ is the entropy produced along a trajectory in
phase space of a microscopically reversible chaotic system satisfying a criterion called the ``Chaotic Hypothesis''. In \cite{GC1995},
 Eq.(\ref{eq-generalFR}) is a theorem in which 
$\Sigma=1$ and there are no adjustable parameters. Attempts to probe the validity of these hypotheses by
experiments and numerical simulations have struggled with the fact
that it is difficult to find systems for which the assumptions of the theorem, are under control.  See \cite{BGC1997} for one such experimental test and further discussion.

Notwithstanding this difficulty, several experimental and numerical studies
have studied the analogue of Eq.(\ref{eq-generalFR})  in
situations where the result may not be expected to hold a-priori. See ~\cite{BCG2008}
and the references therein.  Despite the fact that, in
almost all cases, 
these systems were dissipative and follow microscopically irreversible dynamics, 
they have succeeded in yielding an empirical FR.  Here,
empirical means that the parameter $\Sigma$, which is typically not 1, cannot be 
deduced a-priori from known properties of the system (with the possible exception of the granular experiment discussed in \cite{PVBTW2005}). Not
all of these experiments have provided equally good tests of the FR. In some experiments the fluctuations were Gaussian, in which case Eq.(\ref{eq-generalFR}) is immediately upheld. While it may not be trivial to prove rigorously the Gaussianity of fluctuations in an interacting system, Eq.(\ref{eq-generalFR}) is considered to be most interesting when the PDF of $X_\tau$ is non-Gaussian. Other experiments could not access sufficiently large
 negative fluctuations to provide a statistically convincing test. Efforts are ongoing to try to find good  systems on which to explore the seeming ubiquity of the FR.

From the theoretical perspective, it has been pointed out that any Langevin 
system exhibits a fluctuation relation for a properly defined entropy 
production ~\cite{CCJ2006} (although such a definition need not be unique). 
While one could 
argue that some of the aforementioned experiments could be sensibly modeled by 
Langevin dynamics, most of them measure work or power rather than entropy 
production and the relation between the two is not always obvious. Indeed it 
has been shown analytically ~\cite{FAR2002,FAR2004} that power injection fluctuations in driven 
Langevin systems generally do not exhibit a Fluctuation Relation.  It has also 
been suggested ~\cite{AFMP2001}  that the FR follows
as a property of large deviations having generic Kramer functions, that is,  
those which are differentiable near zero.

In this letter, we contribute to this ongoing discussion in two ways. Firstly, we 
present results for a new system which exhibits an empirical FR - local power fluctuations in turbulence. While it may not help much to simply add another system to the list discussed above, we argue that this system offers considerable advantages as a model system compared with some of the others. It is easy to realise experimentally and numerically, has strongly non-gaussian fluctuations and exhibits abundant negative fluctuations. Furthermore, we deduce an expression for the entropy production rate from known properties of the power distribution that agrees very well with the empirical results.

Secondly, we provide an analytic model of the PDF of the power fluctuations  themselves (before time averaging) based on the PDF products of normal variables. This PDF and its asymptotic properties match very  well with our empirical PDFs and will find many applications beyond the present study. It is worth remembering amid the current flurry of interest in
FRs, that the PDF of the injected power itself is probably of more
direct physical and practical interest than the PDF of its 
time-averaged counterpart. On this basis, the results we present on  the
structure of the power PDF itself are of more general interest, beyond
the topic of FRs. The layout of the paper is as follows. We
first provide a brief discussion of two-dimensional turbulence
and  explain the power measurements in the Eulerian and Lagrangian frames.  We then 
describe the structure and asymptotics of the PDFs of the power and 
compare with the empirical PDFs obtained from our data. Finally we present data demonstrating an empirical FR and explain the original of the value obtained for $\Sigma$.

\section{Two-dimensional Turbulence}
The considerable differences between the physics of turbulence in 2D compared to
3D can be traced to the existence of an additional inviscid invariant (in 
addition to the kinetic energy), known as enstrophy. 
As a result of the dual conservation of energy and
enstrophy in 2D, there are two distinct scaling regimes in 2D turbulence in
the inviscid limit. If energy and enstrophy are injected into the flow at a 
given scale, $l_f$, the enstrophy tends to be transferred to scales smaller than
$l_f$. This process is refered to as a direct cascade. The energy,
on the other hand, tends to be transferred to scales larger than $l_f$. 
This phenomenon, referred to as an inverse cascade is specific to 2D.
According to the classical theory of Kraichnan ~\cite{KRA1967}, the energy spectrum in the
direct cascade regime scales as $k^{-3}$ where $k$ is the modulus of the
Fourier space wave-vector whereas the energy spectrum in the inverse cascade
regime scales as $k^{-5/3}$. For detailed  review of the theory and 
phenomenology of 2D turbulence see ~\cite{LES1997} and the references therein.

Although we have studied the statistics of the energy injection 
rate in both the direct and inverse cascade regimes, in this letter we primarily discuss results for the inverse cascade regime. The analysis performed for the direct cascade regime was observed to be essentially similar to the inverse cascade results.
 The only related study which we are aware of is ~\cite{LPF1996} where the statistics of 
 the global (i.e. averaged over the entire system) power have been studied experimentally 
 in 3D. Here, we study the local power in 2D as we now explain.The instantaneous 
 rate of energy input at a point, $\bf{x}$ in the fluid is given by the scalar product of the 
instantaneous fluid velocity at $\bf{x}$, $\bf{v}(\bf{x},t)$, and the instantaneous
body force, $\bf{f}(\bf{x},t)$, acting on the fluid at $\bf{x}$. 
To construct the PDF of $\cdot\bf{f}$ there are two approaches. One can fix the time and measure the $\bf{v}$ and $\bf f$ at different points and construct the PDF over the values of the local power at different spatial points. This is called an Eulerian measurement. Alternatively we may follow the trajectory, 
$\bf{x}(t)$, of a particular fluid element as it is advected by the turbulence and measure the force, $\bf{f}(t) = \bf{f}(\bf{x}(t),t)$, velocity, $\bf{v}(t)=\bf{v}(\bf{x}(t),t)$ and power $p(t)=\bf{v}(t)\cdot\bf{f}(t)$ experienced by this element. We then construct the PDF over the values of the power at different times. This is called a Lagrangian measurement. We have measured the power distribution in both the Eulerian and Lagrangian frames but focus more on the latter. The variance of the velocity is slighly larger in the Lagrangian frame but once this has been taken into account, the power statistics look very similar for both cases. For a careful discussion of the similarities and differences between the Eulerian and Lagrangian frames in 2D turbulence see \cite{RE2007}.

The forcing used in our simulations, models
that used to excite turbulence in electromagnetically driven 
fluid layers, a popular method of generating 2D turbulence in the laboratory
~\cite{RDE2005}.
We simulated a biperiodic domain of size $2\pi$. For details of the simulations
see ~\cite{BC2007b}.  Aside from the issue of experimental relevance, the reason
to choose the electromagnetic forcing protocol over the stochastic forcing 
protocols traditional for numerical simulations of turbulence is that we 
do not wish to introduce an additional stochasticity into the system.
All of the power fluctuations should come from the intrinsic stochasticity of 
the flow. To generate a direct cascade, we forced at large scales using a distorted magnetic field perpendicular to the layer with characteristic wave number 5. To generate an inverse cascade, we forced at smaller scales
with a magnetic field having characteristic wavenumber 35.  For further 
details of the turbulence diagnostics and
spectra see ~\cite{BC2007b}. The fluid was driven with a direct current in the 
$x$-direction. The only component of the force is then in the $y$ direction so 
that $p(t)$ is just given by a simple product of y-components of force and 
velocity rather than a scalar product. Results for alternating current are
discussed in ~\cite{BC2007b}. One should appreciate that although the forcing is deterministic (the disorder of the underlying magnetic field is quenched), in the Lagrangian frame, it is sampled by a random trajectory so that the force acting on a Lagrangian particle is effectively stochastic.  Similarly, from an Eulerian perspective, the force is a random variable in the sense that it takes a distribution of values at different spatial points.

\begin{figure}
\begin{center}
  \includegraphics[width=6.5cm]{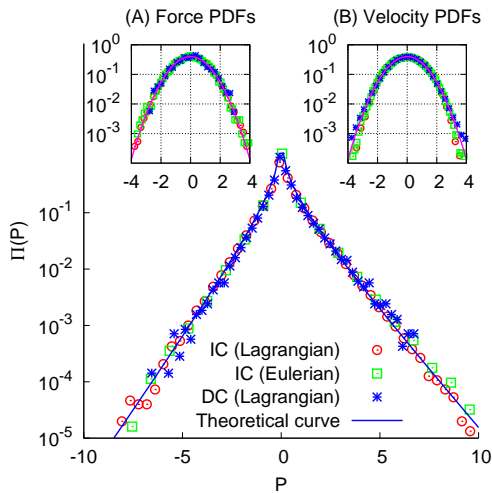}
\end{center}
\caption{PDFs of the power normalized by the product of standard deviations of force ($\sigma_f$) and velocity ($\sigma_v$) for both the Inverse (circles) and Direct cascade (asterisks) regimes in the Lagrangian frame as well as the Inverse Cascade regime in the Eulerian frame (squares) . The solid line is the asymptotic behavior predicted by Craig's XY-model in Eq. (\ref{eq-PxyTails}) for the power distribution. The insets show the corresponding (A) force PDFs  and (B) velocity PDFs for the Inverse and Direct cascade regimes in Lagrangian framKF2007e, the Inverse cascade regime in Eulerian frame and their respective Gaussian fits. Please note the force and velocity PDFs have been normalized by their respective standard deviations. }
\label{fig-forceVelocityPDFs}
\end{figure}

Fig. \ref{fig-forceVelocityPDFs} shows plots of the PDFs of the dimensionless power for different regimes. The non-dimensionalisation has been done by normalising with the standard deviations of the force and velocity, $P = p/(\sigma_v \sigma_f)$. This normalisation quantifies our earlier assurances that the choice of inverse versus direct cascade or Eulerian versus Lagrangian frame is unimportant. The shape of the power distribution is practically the same once the data have been rescaled by these standard deviations. We should point out at this point that very similar PDFs have been observed in similar contexts \cite{FM2004,CEERWX2006,FLF2008}. It turns out that the Lagrangian force and velocity are quite close to Gaussian in all cases (see also \cite{KF2007}). The PDFs, again rescaled by the appropriate standard deviations, are shown in  the insets for Fig.~\ref{fig-forceVelocityPDFs} along with a Gaussian indicator curve. The
measured standard deviations and instantaneous correlation coefficient between $v(t)$ and $f(t)$ for the inverse and direct cascades are given in Table~\ref{tab-data}

\begin{table}
\caption{Statistical parameters of the Lagrangian force and velocity.}
\label{tab-data}
\begin{center}
\begin{tabular}{|l|l|l|l|}
\hline
Simulation & $\sigma_f^2$ & $\sigma_v^2$ & $\rho$ \\
\hline
Inverse cascade & 0.082 & 0.047 & 0.110 \\
Direct cascade & 0.338 & 0.589 & 0.144 \\
\hline
\end{tabular}
\end{center}
\end{table}

\section{Local power and the Fluctuation Relation}

The power PDF, is strongly
non-gaussian as can be seen from Fig.~\ref{fig-forceVelocityPDFs}. For both the
direct and inverse cascades, it is strongly cusped at 0 with asymmetric
exponential tails. Given that $p(t) = v_y(t)\, f_y(t)$ and both $v_y(t)$ and
$f_y(t)$ are close to Gaussian, one may ask if the non-trivial 
features of the power distribution can be understood by considering the 
distribution of the product of two correlated normal variables ~\cite{BC2007b}
\footnote{Describing the PDF of injected power using two correlated normal
variables has been proposed independently in the context of wave turbulence by E. Falcon, S. Auma\^{i}tre, C. Falc\'{o}n, C. Laroche and S. Fauve.\cite{FLF2008}}. The 
distribution of the 
product, $P=v f$, of two joint normally distributed variables, $v$
and $f$, with zero mean, standard
deviations $\sigma_f$ and $\sigma_v$ and correlation coefficient $\rho$ was
considered by Craig ~\cite{CRA1937}. The PDF of the product is:
\begin{equation}
\label{eq-Pxy}
\Pi_{\rm xy}(P) = \frac{\sqrt{\Lambda^{+}\Lambda^{-}}}{2 \pi} \int_{-\infty}^{\infty} \frac{e^{i P w} d w}{\sqrt{(w - i \Lambda^{+})(w + i \Lambda^{-})}}.
\end{equation}
where $\Lambda^\pm = (\sigma_x\sigma_y(1\pm\rho))^{-1}$. In general, this
integral cannot be further simplified. The asymptotic
behaviour, however, is relatively easily calculated. We find that the tails
are indeed almost exponential with the positive tail decaying more
slowly than the negative tail. Full details are given in ~\cite{BC2007b}:
\begin{equation}
\label{eq-PxyTails}
\Pi_{\rm xy}(P) \sim \left\{
\begin{array}{ll}
\sqrt{\frac{\Lambda^{+}\Lambda^{-}}{\pi(\Lambda^{+}+\Lambda^{-})}} \frac{{\rm e}^{-\Lambda^{+} P}}{\sqrt{P}}& \mbox{$p>0$}\\
\sqrt{\frac{\Lambda^{+}\Lambda^{-}}{\pi(\Lambda^{+}+\Lambda^{-})}} \frac{{\rm e}^{-\Lambda^{-} \left|P\right|}}{\sqrt{\left|P\right|}}& \mbox{$p<0$}
\end{array} \right.
\end{equation}
Fig.\ref{fig-forceVelocityPDFs} super-imposes this asymptotic behaviour on the
empirical PDFs for the measured values of $\sigma_f$, $\sigma_v$ and $\rho$
for each cascade regime tabulated in Table \ref{tab-data}. Note that we plot the normalized power $P = p/(\sigma_v \sigma_f)$ for the asymptotic behavior predicted by Eq. (\ref{eq-PxyTails}). The agreement is excellent. This agreement is a result of the fact that the values of $\rho$ are close for these situations. Other values of $\rho$ for other systems may lead to a greater or lesser degree of asymmetry. Near zero, where the asymptotic expression in Eq.
(\ref{eq-PxyTails}) is not valid, Eq.(\ref{eq-Pxy}) can be shown to
be logarithmically singular ~\cite{BC2007b}. Although our data is insufficient 
to resolve such a weak divergence, this explains the cusp-like structure
of the empirical PDFs near zero. We conclude from this analysis that most of
the interesting features of the power PDF can be understood in statistical
terms. Dynamical information enters only through the values of  $\sigma_f$, 
$\sigma_v$ and $\rho$ which, we should emphasise, are measured rather than
calculated from first principles.

\begin{figure}
\begin{center}
  \includegraphics[width=6.5cm]{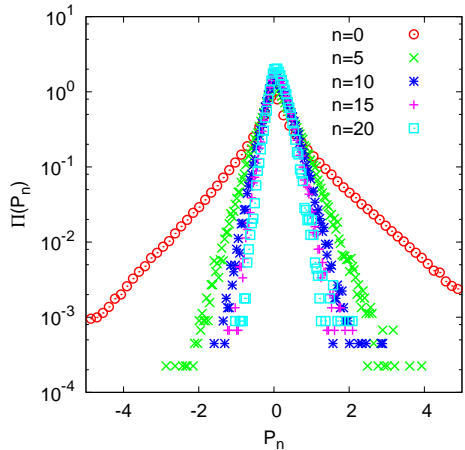}
\end{center}
\caption{PDFs of the coarse-grained dimensionless power ($P_n$) as defined in Eq. (\ref{nondimp}) for the Inverse cascade regime. The PDFs are plotted for the dimensionless averaging time $n = \tau/\tau_c$ = 0, 5, 10, 15, and 20 respectively. The coarse-graining leads to narrowing of the PDFs with increasing n as expected. However they retain the non-gaussian features of the non-coarsegrained distribution.}
\label{fig-coarsegrainedPDFs}
\end{figure}

\begin{figure}
\begin{center}
  \includegraphics[width=6.5cm]{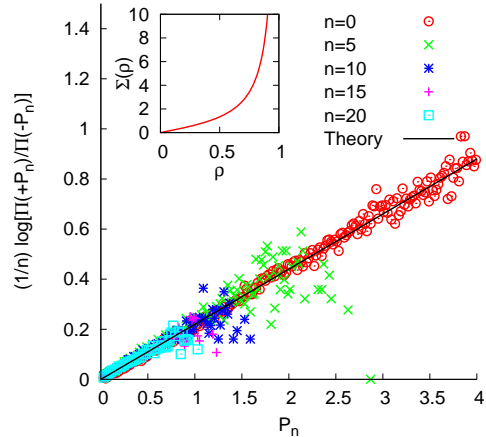}
\end{center}
\caption{Plots for the Left Hand Side of Eq. (\ref{eq-dataCollapse}) for different coarse-graining intervals $n = $ 0, 5, 10, 15, and 20. The solid line is the theoretical prediction for $\Sigma$ deduced directly by applying the XY-model to the non-coarsegrained power distribution. As is evident, the theoretical prediction for $\Sigma$ is in excellent agreement with the observed empirical Fluctuation Relation. The inset shows the functional dependence of the entropy rate $\Sigma$ to the correlation coefficient $\rho$ which essentially decides the asymmetry of the power distribution.}
\label{fig-FR}
\end{figure}

We note that Eq.(\ref{eq-PxyTails}) shows that the tails of the power PDF
satisfy a fluctuation relation of the form Eq.(\ref{eq-generalFR}). Taking the ratio of the right to the left tail from Eq. (\ref{eq-PxyTails}) results in an expression for $\Sigma$:

\begin{equation}
\label{Sigma}
\Sigma = \frac{2\rho}{(1-\rho^2)}
\end{equation}

The entropy rate $\Sigma$ is now expressed purely in terms of the correlation coefficient $\rho$ , which is known to decide the degree of the asymmetry of the power distribution, see ~\cite{BC2007b}. It is
natural to expect that the time integrated power should satisfy a similar relation with a comparable value of $\Sigma$.

We test the FR for the Lagrangian power in the inverse cascade regime since this is the regime for which we have the most data. We coarsegrained the measured power over time windows which are longer than the correlation time, $\tau_c$, of the Lagrangian power signal. These 
correlation times are quite short since Lagrangian quantities tend to 
decorrelate very quickly. For the inverse cascade we measured $\tau_c = 0.144$ and
for the direct cascade, $\tau_c = 0.084$. These compare to large eddy turnover times of about $22$ and $10$ respectively. We then define the non-dimensionalised 
coarse-grained power over a non-dimensional time-interval of length $n = \tau/\tau_c$:
\begin{equation}
\label{nondimp}
P_n(t) = \frac{1}{P_0} \frac{1}{\tau} \int_t^{t+\tau} p(t^\prime)\, dt^\prime.
\end{equation}
Here, $t^\prime$ is a dimensionless time normalised by the correlation time $\tau_c$. In fluctuation relation analyses, it is customary to normalize the signal of interest by its long-time average. However, in the present case, we normalise the power $p$ by $P_0 = \sigma_v \sigma_f$. From Eq. \ref{eq-Pxy} we see that this is an attractive normalisation since it reduces the underlying PDF to a function of $\rho$ only.
The statistics of $p_n$ are stationary in $t$ since the underlying turbulence
had reached a stationary state before we started to gather power statistics. 
The coarse-grained PDFs are shown in Fig.~\ref{fig-coarsegrainedPDFs} for
dimensionless averaging times ranging from $n = \tau/\tau_c =$ 0 to 20. They clearly narrow
as one integrates out more and more of the large fluctuations but retain their
non-gaussian character as the averaging proceeds. If the FR is satisfied, then
Eq(\ref{eq-generalFR}) suggests that all the data should collapse 
according to the following formula:
\begin{equation}
\label{eq-dataCollapse}
\frac{1}{n} \log \left[ \frac{\Pi(+p_n)}{\Pi(-p_n)}\right]=  \Sigma\,p_n.
\end{equation}
Additionally, if the asymmetry of the power distribution observed for the non-coarsegrained power PDF is retained after coarsegraining, one expects the left hand side of eq. (\ref{eq-dataCollapse}) should scale with the slope predicted by eq. (\ref{Sigma}). This collapse of the data is demonstrated in Fig~\ref{fig-FR} for the Inverse cascade. The value of  $\Sigma$ obtained from Eq. (\ref{Sigma}) after substituting the appropriate value for $\rho$ from Table \ref{tab-data} is 0.22. The corresponding value of $\Sigma$ obtained from the empirical FR plot in Fig. \ref{fig-FR} is 0.217 for the inverse cascade regime. The solid line in Fig \ref{fig-FR} is the theoretical prediction following eq. \ref{Sigma} and is in excellent agreement with the empirical Fluctuation Relation. Note that owing to the coarsegrained power and time being non-dimensional, the entropy rate $\Sigma$,  is also a dimensionless quantity. In the inset to Fig. \ref{fig-FR} we plot the expected functional dependence of $\Sigma$ with respect to the correlation coefficient $\rho$. The value of $\rho$ and hence the asymmetry of the power PDF changes with the strength of the turbulence, which is normally quantified in terms of the dimensionless Reynolds number ($Re$). Depending upon how $\rho$ varies with respect to $Re$, one is then in a position to understand how the entropy rate $\Sigma$ varies as a function of the Reynolds number. Such an analysis is however beyond the scope of the present work.

\section{Summary}
In conclusion, we have demonstrated an empirical FR for the statistics of
Lagrangian power in turbulence. It is an attractive system in which to study
non-equilibrium fluctuations since the fluctuations are strongly non-Gaussian
with an abundance of strong negative fluctuations. In addition, we have
phenomenologically established the analytic form of the  underlying power
PDF. Its qualitative features can be understood in terms of the statistics of
products of normally distributed variables.  Most importantly, we are able to arrive at a theoretical expression for $\Sigma$ that is in excellent agreement with the empirical results.

\acknowledgments
This work was carried out under the auspices of the National Nuclear Security
Administration of the U.S. Department of Energy at Los Alamos National
Laboratory under Contract No.  DE-AC52-06NA25396. We acknowledge
helpful discussions with M. Chertkov, R. Ecke, M. Rivera, R. Teodorescu and O. Zaboronski.


\end{document}